\documentclass{egpubl}
\usepackage{sca2020}

\SpecialIssuePaper         

\usepackage[T1]{fontenc}
\usepackage{dfadobe}  

\usepackage{cite}  
\BibtexOrBiblatex
\electronicVersion
\PrintedOrElectronic
\ifpdf \usepackage[pdftex]{graphicx} \pdfcompresslevel=9
\else \usepackage[dvips]{graphicx} \fi

\usepackage{egweblnk} 

\usepackage{amsmath}
\usepackage{amssymb}
\usepackage{dsfont}
\usepackage{booktabs}

\newcommand{\norm}[1]{\|#1\|}
\newcommand{\minus}{\scalebox{0.75}[1.0]{$-$}}

\renewcommand{\paragraph}[1]{\noindent\textbf{#1}.}

\title[ALLSTEPS]%
      {ALLSTEPS: Curriculum-driven Learning of Stepping Stone Skills}

\newcommand{\spsp}{\hspace*{0.2cm}}
\author[Zhaoming Xie \& Hung Yu Ling \& Nam Hee Kim \& Michiel van de Panne]
{\parbox{\textwidth}{\centering Zhaoming Xie\thanks{These authors contributed equally to this work.}
        \spsp Hung Yu Ling$^\fnsymbol{footnote}$
        \spsp Nam Hee Kim
        \spsp Michiel van de Panne
        }
        \\
{\parbox{\textwidth}{\centering University of British Columbia, Canada\\
      }
}
}

\begin{document}

\teaser{
  \includegraphics[width=0.9\linewidth]{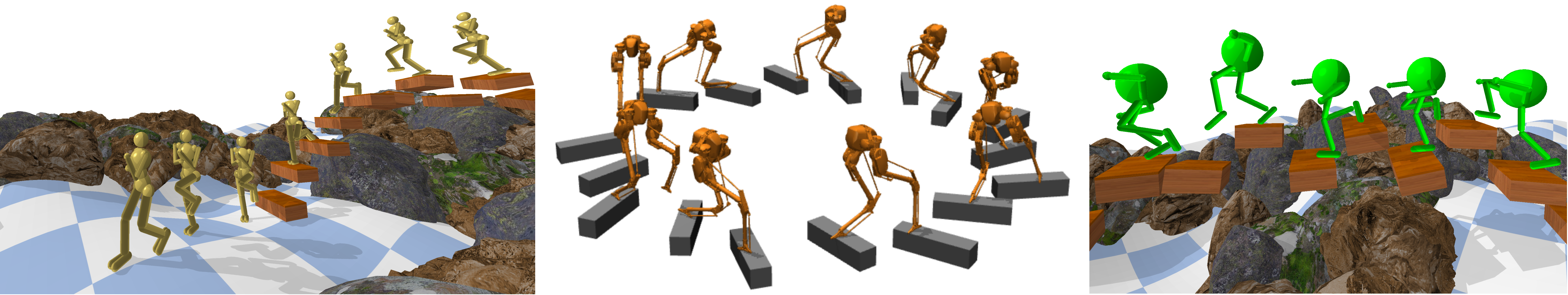}
  \centering
  \caption{Humanoid (left), Cassie (middle), and Monster (right) walk across randomly generated stepping-stone terrain.}
\label{fig:teaser}
}

\maketitle

\begin{abstract}

Humans are highly adept at walking in environments with foot placement constraints,
including stepping-stone scenarios where footstep locations are fully constrained.
Finding good solutions to stepping-stone locomotion is a longstanding and fundamental challenge for animation and robotics.
We present fully learned solutions to this difficult problem using reinforcement learning. 
We demonstrate the importance of a curriculum for efficient learning and evaluate four possible curriculum choices compared to a non-curriculum baseline.
Results are presented for a simulated humanoid, a realistic bipedal robot simulation and a monster character, 
in each case producing robust, plausible motions for challenging stepping stone sequences and terrains.


\end{abstract}  
\section{Introduction}

Bipedal locomotion is a fundamental problem in computer animation and robotics, and there exist
many proposed data-driven or physics-based solutions.  
However, a principal raison-d'être for legged locomotion is the ability to navigate over challenging irregular terrain, and this is unfortunately not reflected in the bulk of locomotion work, which targets flat terrain.
Traversing irregular terrain is challenging, with the limiting case being that of navigation across sequences of stepping-stones that fully constrain each footstep. 
We wish to learn physics-based solutions to this classical stepping stone problem from scratch, 
i.e., without the use of motion capture data,. 
The limits of the learned skills should ideally stem from the physical capabilities
of the characters, and not from the learned control strategy.

We investigate the use of deep reinforcement learning (DRL) for computing solutions to this problem.
We find a curriculum-based solution
to be essential to achieving good results; the curriculum begins with easy steps and advances to challenging steps.
We evaluate four different curricula, which each advance the learning based on different principles, 
and compare them against a no-curriculum baseline.
Challenging stepping stone skills are demonstrated on a humanoid model, a fully-calibrated simulation of a large bipedal robot and a monster model.
Finally, we demonstrate that the stepping stone policies can be directly applied to walking on 
challenging continuous terrain with pre-planned foot placements.

\noindent Our contributions are as follows:
\begin{itemize}
\item We show how control policies for challenging stepping stone problems can be learned from scratch
using reinforcement learning, as demonstrated across 3 bipeds and 2 simulators.
Leveraging careful reward design, we are to learn control policies producing plausible motions, without the use of reference motion data.
\item We demonstrate the critical role of a curriculum to circumvent local minima in optimization and that support efficient learning for this task.
We evaluate four curricula in comparison to a no-curriculum baseline.
\item We demonstrate that the stepping stone control policies are directly transferable to locomotion on continuous terrain. 
The learned stepping stone skills thus serve as a general solution for navigating many types of terrain.
\end{itemize}

\section{Related Work}
\label{sec:related}

The stepping stone problem is of interest to many, including: animation and robotics, as will be discussed
in more detail below, gait and posture, e.g., \cite{lindemann2013assessment,potocanac2014response},
and neuromotor control, e.g., \cite{patla2003far,matthis2014visual}.
In what follows below, we focus principally on related work in animation and robotics.

\subsection{Learning Bipedal Locomotion} 
Considerable progress has been made towards learning control policies for locomotion in the context of physics-based 
character animation,
often via deep reinforcement learning. In many cases, these aim to satisfy an imitation objective
and target motions on flat terrain,
e.g.~\cite{liu2016guided,Peng_2017_deeploco,liu2017learning,Park_2019_siggraph_asia_learn_and_predict,bergamin2019drecon}.
Other solutions learn in the absence of motion capture data, also for flat terrain,
e.g.,~\cite{Yu_2018_Symmetry,ScalableMuscle,jiang2019biologically_real}.
Environment information such as height maps \cite{peng2016terrain,peng2018deepmimic} or egocentric vision
\cite{merel2018hierarchical} can be fed into the policy to adapt to some degree of terrain
irregularity. Learned kinematic locomotion controllers have recently achieved impressive results for
terrains that includes hills and obstacles \cite{holden2017phase_neural_net,MANN}, although equivalent capability
has not been demonstrated for physically simulated characters. The stepping-stone problem has also been
tackled using trajectory optimization, e.g., \cite{safonova2004synthesizing_low_dim_motion}.

\subsection{Walking on Stepping Stones}
Precise foot placement is needed to achieve stepping stone capability. There are many works in the
robotics literature that achieve this capability by utilizing path planning techniques, including mixed integer programming \cite{Debits_2014_Humanoid} or variants of A* search
\cite{chestnutt2005footstep,ihmc_2019_humanoid}. 
Such techniques are often limited to flat terrain \cite{chestnutt2005footstep} 
or to slow motions informed by quasi-static solutions~\cite{tonneau2018efficient}. 
Another line of work uses a gait library \cite{Nguyen_2017_stepping_stones}, consisting of trajectories for different steps that are computed offline and are then used to achieve stepping stone walking on a bipedal robot whose motion is restricted to the sagittal plane.

3D stepping stones capability has been shown in several simulated bipedal
character models. \cite{Nguyen_2016_3D_stepping_stones} approach this via the use of control barrier
function, although that heavily relies on the feasibility of the resulting Quadratic Programming
problem, which is not always satisfied. Furthermore, while the simulated model is 3D, the steps
themselves are placed in a straight line on a horizontal plane, i.e., only have distance variation, and thus 
no height variation or turning. 
There are also works in computer animation literature demonstrating 3D stepping skills, e.g. \cite{Coros_2008_constrained} and \cite{mordatch2010robust}, generally with limited realism and capabilities.
Terrain-adaptive motion has been tackled using end-effector trajectory planning and 
quadratic programming to produce controllers for navigating on uneven terrains, such as stairs 
and hurdles~\cite{TerrainAdaptive2010}. This uses manually-crafted representations and motion cycle phases.
Our work differs in terms of its focus on the potential of learning-based approaches and on high-difficulty stepping scenarios.
Lastly, foot placement has also been used as guidance for reinforcement learning algorithms to achieve path following capability for a simulated biped \cite{Peng_2017_deeploco}. 
It is used to parameterize the possible steps, on flat terrain, and in practice it does not always do well at reaching the desired foot placements.

\subsection{Curriculum-based Learning} 
Curriculum learning is the learning process where task difficulty is increased overtime during
training \cite{2009-ICML-curriculum}. It has also been applied for synthesizing navigation or
locomotion policies,
e.g. \cite{yin2008continuation,karpathy2012curriculum,florensa2017reverse_curriculum,heess2017parkour,Yu_2018_Symmetry}.
The task difficulty is usually determined by human intuition. Teacher-student curriculum learning
\cite{matiisen2019teacher_student_curriculum} uses the progress on tasks as a metrics for choosing
the next tasks, and demonstrates automatic curriculum choices on decimal number additions of different
lengths and Minecraft games. Intrinsic motivation \cite{forestier2017intrinsic_curriculum} can also
help to let robots begin from simple goals and advance towards complicated goals. A curriculum policy
\cite{narvekar2019curriculum_policy} can further be learned by formulating the curriculum learning
process as a Markov Decision Process (MDP). More recently, \cite{wang2019poet} proposes the POET
algorithm that allows a 2D walker to solve increasingly challenging terrains by co-evolving the environment and the policy. 
Reverse curriculum learning has been shown to be effective at balancing uneven data generation in DRL. For example, \cite{Won_2019_siggraph_asia_body_shape} and \cite{Park_2019_siggraph_asia_learn_and_predict} propose a form of adaptive sampling where more difficult tasks are given higher priority during training.

\section{System Overview}

\begin{figure}
  \centering
  \includegraphics[width=0.9\columnwidth]{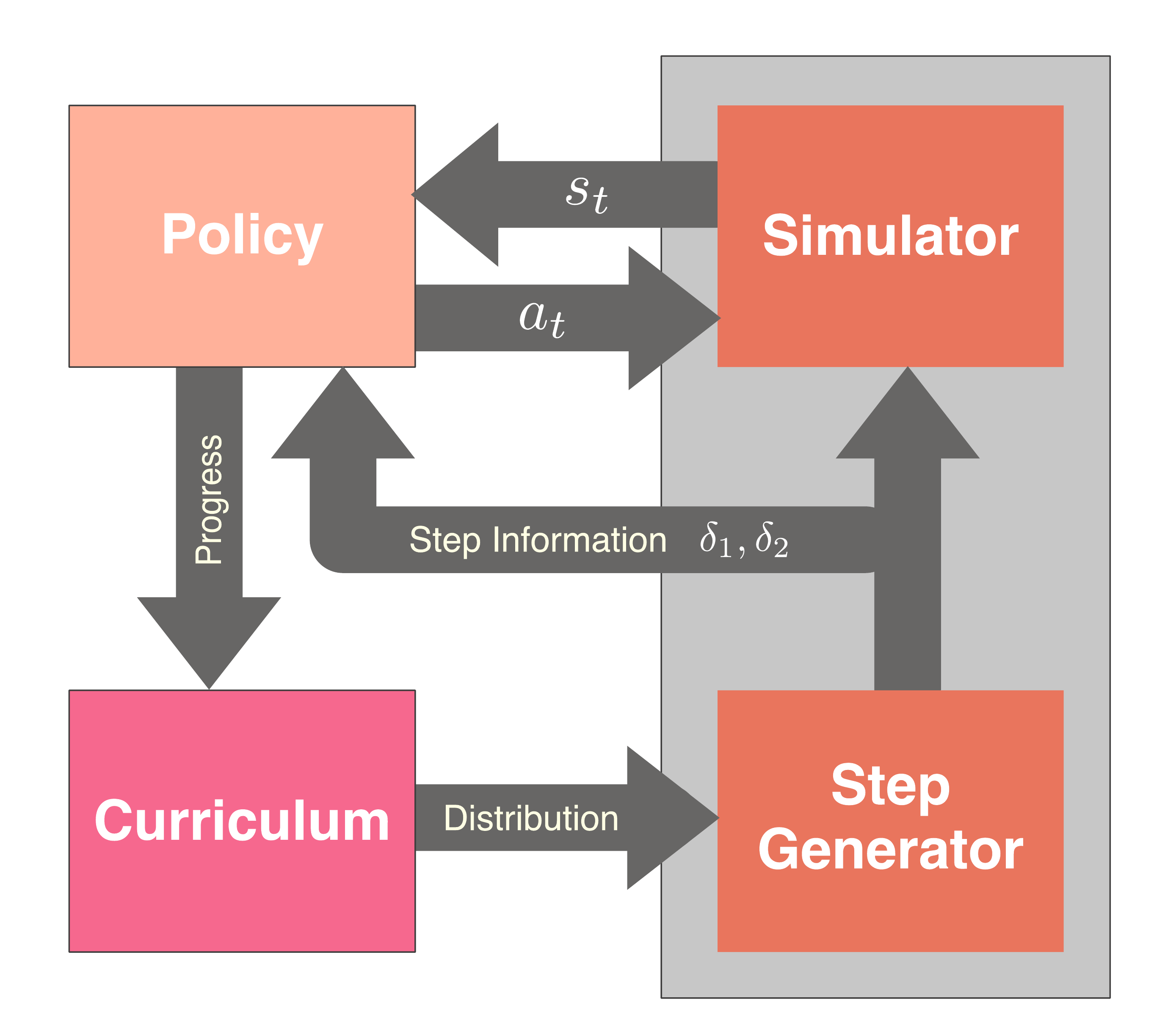}
  \caption{Overview of our curriculum learning system.  The curriculum module improves learning efficiency by dynamically adjusting the terrain difficulty according to the progress of the policy.}
  \label{fig:system-overview}
\end{figure}

An overview of our system is shown in Figure~\ref{fig:system-overview}. The environment consists of a physics simulator and a step generator. 
The step generator samples a random sequence of steps from a given probability distribution of the step parameter space. 
In the case where no curriculum is applied, the step distribution is uniform across the parameter space for the entire duration of training.
In contrast, a curriculum dynamically adjusts the step distribution depending on the progress made by the policy.
We experiment with four different curricula and a baseline, each having its own motivation and benefits.
We show experimentally that curriculum learning, when applied appropriately, allows the policy to solve the stepping stone task, which is otherwise very challenging with standard reinforcement learning.

The remaining of the paper is organized as follows: stepping stones task definition and character modelling (\S\,\ref{sec:simulation-environments}), reinforcement learning and reward specifications (\S\,\ref{sec:learning-and-reward}), learning curricula (\S\,\ref{sec:curriculum}), experimental results (\S\,\ref{sec:results}), and discussions (\S\,\ref{sec:discussion}).
\section{Simulation Environments}
\label{sec:simulation-environments}

We now describe the stepping stones parameter space and character models.
We experiment with three different characters, Humanoid, Cassie and Monster, 
to show that the proposed curricula provide a robust approach for learning stepping stones skills.

\subsection{Stepping Stones Generation}
\label{sec:steps-generation}

\begin{figure}%
  \centering
  \includegraphics[width=.9\columnwidth]{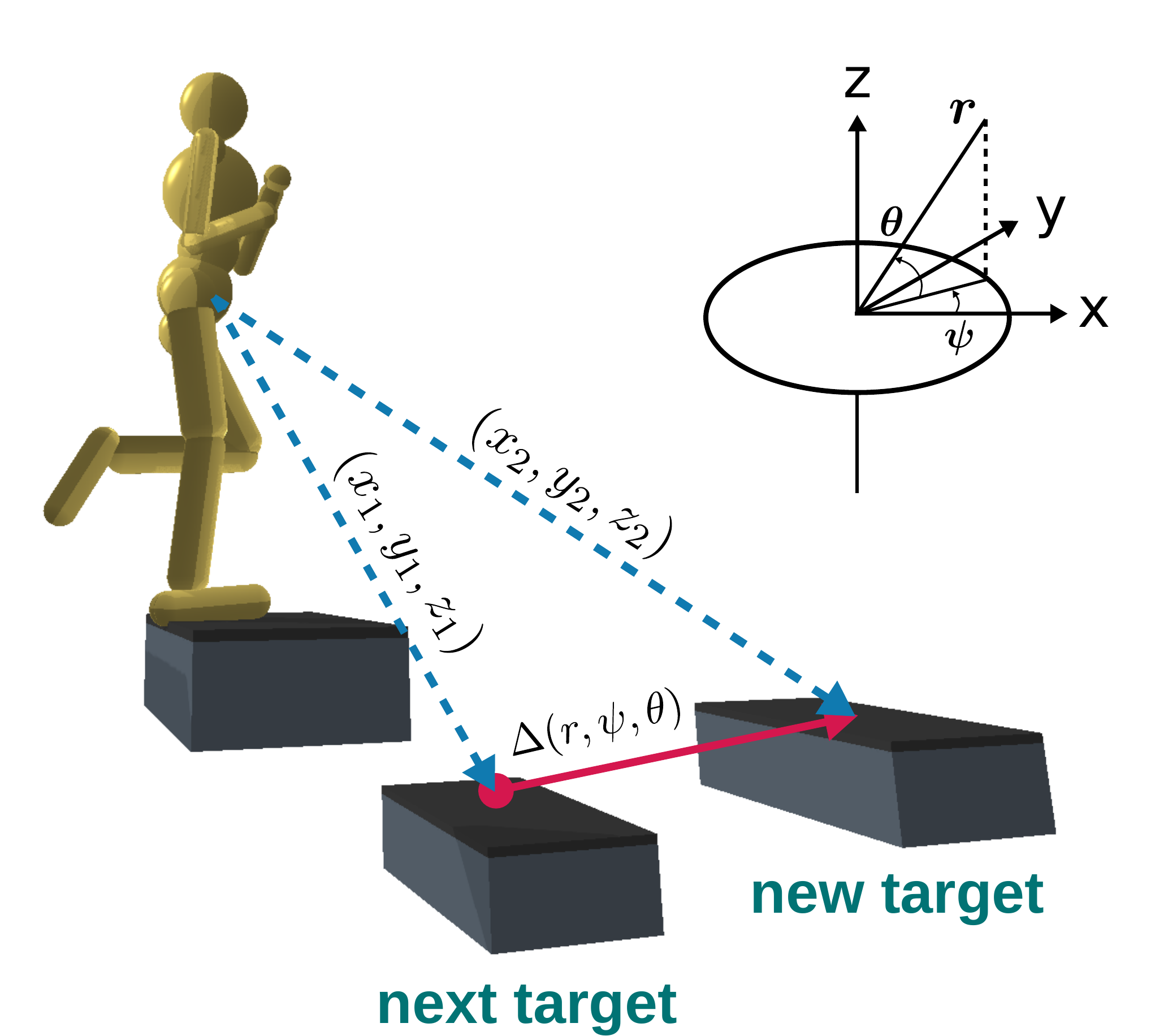}
  \caption{Illustration of the stepping stone problem. The character observes the position of the next two steps with respect to its center-of-mass. The new target is generated from a distribution parameterized by three parameters: $r, \psi$ and $\theta$.}
  \label{fig:step-generation}
\end{figure}

In the stepping stones task, the goal of the character is to make precise foot placements on a sequence of discrete footholds.
The character receives foothold positions of the two upcoming steps in its root space, i.e., $(x_1, y_1, z_1)$ and $(x_2, y_2, z_2)$ as shown in Figure~\ref{fig:step-generation}. 
We use two steps since two-step anticipation yields better performance than a single-step~\cite{Nguyen_2017_stepping_stones}, 
and it has been found that further anticipation may be of limited benefit~\cite{Coros_2008_constrained}.

Successive step placements are generated in spherical coordinates, where the step length $r$, yaw $\psi$, and pitch $\theta$ relative to the previous stone are the controllable parameters. 
This 3D parameter space is also illustrated in Figure~\ref{fig:step-generation}. 
We limit the distance, yaw, and pitch to lie in the intervals 
$[r_{\min}, r_{\max}]$, $[-\psi_{\max}, \psi_{\max}]$, and $[-\theta_{\max}, \theta_{\max}]$ respectively.
During training, we set $\psi_{\max} = 20^{\circ}$ and $\theta_{\max} = 50^{\circ}$, 
which we find experimentally to be the upper limits of our character's capability.
For our 2D step-parameter tests, step distance is sampled uniformly from $[0.65, 0.8]$ meters for the humanoid and $[0.35, 0.45]$ meters for Cassie to account for the differences in character morphology.
A 5D parameter space includes additional roll and pitch variations of the step surfaces,
which supports transfer of the skills to smoothly-varying terrains.
The roll and pitch variation of a step, $(\phi_x, \phi_y)$, is generated by first applying the $\psi$ rotation relative to the previous foothold, then subsequently applying the $\phi_x$ and $\phi_y$ rotations about its x-axis and y-axis.
In effect, this causes the step to become tilted as shown in Figure~\ref{fig:tilted-5d-steps}.

When the character successfully steps on the current target, its position $(x_0, y_0, z_0)$ is immediately replaced by that of the next target $(x_y, y_1, z_1)$, and new target $(x_2, y_2, z_2)$ pops into view.
We introduce an artificial look-ahead delay to allow the stance foot to settle (see Table~\ref{tab:character-summary}), 
by postponing this replacement process for a fixed number of frames.
In practice, the look-ahead delay impacts the speed at which the character moves through the stepping stones 
and also enables it to stop on a given step.
Lastly, to ensure that the character begins tackling variable steps from a predictable and feasible state, 
we fix the first three steps of the stepping stones sequence.
Specifically, the first two steps are manually placed slightly below the character's feet, and the third step is always flat and directly ahead.

\subsection{Character Models}
\label{sec:characters}

\begin{figure}%
  \centering
  \includegraphics[width=0.8\columnwidth]{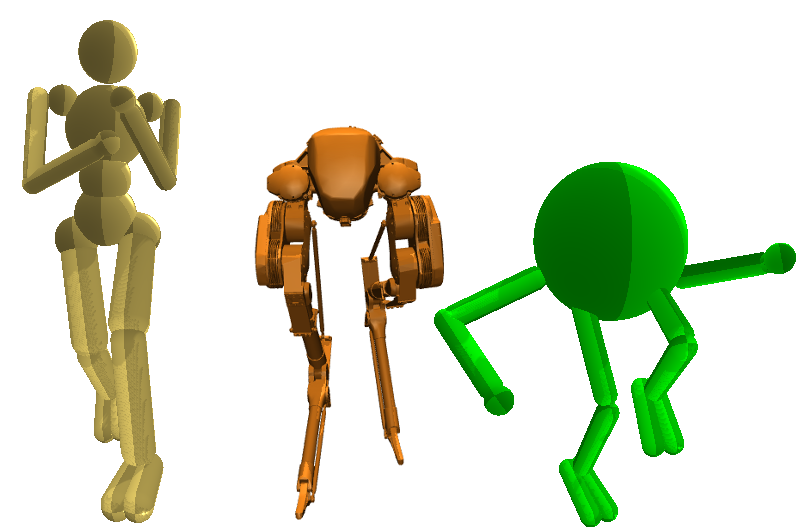}
   \caption{Character models for the Humanoid (left), Cassie (middle), and the Monster (right).}
   \label{fig:characters}
\end{figure}

\begin{table}[t]
\centering
\caption{\label{tab:character-summary}Properties of the characters.}
\begin{tabular}{@{}lrrr@{}}
\toprule
Property & Humanoid & Cassie & Monster\\
\midrule
\textit{Height (m)} & 1.60 & 1.16 & 1.15\\
\textit{Mass (kg)} & 59 & 33 & 33\\
\textit{Action Parameters} & 21 & 10 & 21\\
\textit{Degrees of Freedom} & 27 & 20 & 27\\
\textit{State Features} & 60 & 51 & 60\\
\textit{Maximum Torque (N$\cdot$m)} & 100 & 112.5 & 100\\
\textit{Simulation Freq. (Hz)} & 240 & 1000 & 240\\
\textit{Control Freq. (Hz)} & 60 & 33 & 60\\
\textit{Look-ahead Delay (frames)} & 30 & 3 & 30\\
\bottomrule
\end{tabular}
\end{table}

The character models are shown in Figure~\ref{fig:characters} and the detailed specifications is summarized in Table~\ref{tab:character-summary}.
We focus our experiment and analysis on the Humanoid and Cassie model.
However, we show that the curriculum-based learning pipeline can be directly applied to a third character, the Monster.

\paragraph{Humanoid}
The Humanoid is simulated with 21 hinge joints using PyBullet \cite{pybullet} and is directly torque-controlled.
As is standard in reinforcement learning, we normalize the policy output to be in the interval $[-1, 1]$, then multiply the action value for each joint by its corresponding torque limit.
The state space contains the joint angles and velocities in parent space, roll and pitch of the root orientation in global space, and linear velocities of the body in character root space.
Furthermore, the state space also includes height of the pelvis related to the lowest foot, as well as a binary contact indicator for each foot.
We use the height information to detect when the character falls to early terminate the simulation.
To improve the motion quality, we generate mirrored roll-out trajectories using the DUP method from \cite{2019-MIG-symmetry} to encourage symmetric motions.

Our humanoid character is carefully modelled to reflect joint and torque limits that are close to those documented for humans in \cite{grimmer2015powered_lower_limb}. 
Humanoid characters with unrealistic torque limits often produce unnatural motions unless guided with reference motions, e.g. \cite{peng2018deepmimic, merel2018hierarchical}.
In our experiments, we find, as in \cite{jiang2019biologically_real}, that natural motion is easier to achieve with the use of realistic torque limits.

\paragraph{Cassie}
The action space of Cassie consists of the target joint angles of the ten actuated joints for  
low-level proportional-derivative (PD) controllers.
More concretely, the policy outputs a target joint angle $\alpha_{targ}$ at 33~Hz for each joint, then the joint torque is computed as $\tau = K_P(\alpha_{targ}-\alpha) - K_D \dot{\alpha}$ at 1000~Hz for simulation.
We define the proportional gain $K_P$ for each joint equal to its torque limit, and we set the derivative gain $K_D = K_P / 10$.
The PD controller operates at a much higher frequency than the policy to ensure stability in control and simulation. 

The Cassie model is designed by Agility Robotics, simulated in MuJoCo \cite{mujoco}, and is validated to be very close to the physical robot \cite{2019-CORL-cassie}. 
Designing robust controllers for Cassie is challenging since it has 20-DoF while only having ten actuators.
Furthermore, due to the strong actuators on the robot, it is difficult to obtain high quality motion directly with a simple reward specification. 
To bootstrap stepping stones training, we follow \cite{2018-IROS-cassie} to first obtain a natural forward-walking policy by tracking a reference motion. 
The reference motion contains a walk cycle with a period of 0.84 seconds, or 28 control time steps.
A phase variable is added to the state space to track the timing in a gait cycle. 
In particular, it increases linearly over time from 0 to 1, and resets to 0 at the end of the 28 frames.
Since only a single reference walk cycle is used, the phase variable is purely a function of time, and is not dependent on foot strike timings.
The reference motion is discarded after the base controller is trained, i.e., it is not used during training in the stepping stone environment. 
However, the phase variable is still used as input to the controller, which limits the stepping motion to a fixed period.
Since contact state can be estimated from the phase variable, Cassie does not have the binary contact indicators used in the Humanoid.

\paragraph{Monster}
The third character, the Monster, is identical to the Humanoid except for body morphology, mass distribution, and slightly weaker arms.

\section{Learning Control Policies}
\label{sec:learning-and-reward}

We use reinforcement learning to learn locomotion skills. 
However, as we show in Section~\ref{sec:results}, reinforcement learning alone, without curriculum, is insufficient for solving the stepping stones task.
In this section, we provide the background for actor-critic-based policy-gradient algorithms.
Importantly, the critic module can be used to estimate the performance of the policy, as shown in \cite{Won_2019_siggraph_asia_body_shape, Park_2019_siggraph_asia_learn_and_predict}.
Our adaptive curriculum (\S\,\ref{sec:adaptive_curriculum}) uses the critic to adjust the task difficulty.

\subsection{Proximal Policy Optimization with Actor-Critic}
In reinforcement learning, at each time $t$, the agent interacts with the environment by applying an action ($a_t$) based on its observation ($o_t$) from the environment and receive a reward, $r_t = r(o_t, a_t)$, as feedback. 
Usually the agent acts according to a parametrized policy $\pi_\theta(a | o)$, where $\pi_\theta(a | \cdot)$ is the probability density of $a$ under the current policy. 
In DRL, $\pi$ is a deep neural network with parameters $\theta$. 
The goal is to solve the following optimization problem:
\begin{align*}
\mathop{\mathrm{maximize}}_\theta J_{RL}(\theta) = E_{a_t\sim \pi_\theta(\cdot | o_t)}\left[\sum_{t=0}^{\infty}\gamma^{\;t} r(o_t, a_t) \right],
\end{align*}
where $\gamma \in (0, 1)$ is the discount factor so that the sum converges.

We solve this optimization problem with a policy-gradient actor-critic algorithm, updated using proximal policy optimization (PPO) \cite{schulman2017ppo}. We choose PPO because it is simple to implement and effective for producing high quality locomotion solutions, as demonstrated in previous work, e.g.~\cite{peng2018deepmimic, Yu_2018_Symmetry, Park_2019_siggraph_asia_learn_and_predict, Won_2019_siggraph_asia_body_shape}.

The critic, or the value function, computes the total expected reward a policy can get when starting from an observation $o$.  
The value function is defined for a policy $\pi$ as:
\begin{align*}
V^\pi(o) = E_{o_0=o, a_t\sim \pi(\cdot | o_t)}\left[\sum_{t=0}^{\infty}\gamma^{\;t} r(o_t, a_t) \right].
\end{align*}
In DRL, the total expected reward can often only be estimated, and so we collect trajectories by executing the current policy.
Let an experience tuple be $e_t = (o_t, a_t, o_{t+1}, r_t)$ and a trajectory be $\tau = \{e_0, e_1, \dots, e_T\}$, a Monte Carlo estimate of the value function at $o_t$ can be recursively computed via
$$V^{\pi_{\theta}}(o_t) = \gamma\;V^{\pi_{\theta}}(o_{t+1}) + r_t, $$ 
with $V^{\pi_{\theta}}(o_T) = r_T + \gamma V^{\pi_{\theta_{old}}}(o_{T+1})$. 
The value estimate is used to train a neural network-based value function using supervised learning in PPO.
In policy-gradient algorithms, the value function is usually only used for computing the advantage for training the actor.


The policy, or the actor, is updated by maximizing
\begin{align*}
L_{ppo}(\theta) = \frac{1}{T}\sum_{t=1}^T
 \min(\rho_t\hat{A}_t, \; \text{clip}(\rho_t,1-\epsilon,1+\epsilon)\hat{A}_t),
\end{align*}
where $\rho_t = \pi_{\theta}(a_t | o_t) \mathbin{/} \pi_{\theta_{old}}(a_t | o_t)$ is an importance sampling term used for calculating the expectation under the old policy $\pi_{\theta_{old}}$ and $A_t = V^{\pi_\theta} - V^{\pi_{\theta_{old}}}$ is the advantage estimation. 


\subsection{Reward Design}
\label{sec:rewards}

Despite recent advancements in DRL algorithms, it remains critical to design suitable reward signals to accelerate the learning process. 
We describe the reward specifications used for the stepping stones environment below.

\paragraph{Hitting the Target}
The immediate goal of the character is to place one of its feet on the next stepping target. 
We define the target reward as 
$$r_{target} = k_{target}\exp(\minus d / k_d),$$
where $d$ is the distance between the center of the step target and its contacting foot.
We use $k_{target}$ and $k_d$ to define the magnitude and sensitivity of the target reward. 
To account for the differences in body morphology of the Humanoid and Cassie model, we use $k_{target} = 50$ and $k_d = 0.25$ for the Humanoid and $k_{target} = 20$ and $k_d = 0.1$ for Cassie.
The sensitivity term is chosen to reflect the approximate length of the foot.  
Note that the character receives the target reward only when contact with the desired step is made, otherwise it is set to zero.

In the initial stages of training, when the character makes contact with the target, the contact location may be far away from the center. 
Consequently, the gradient with respect to the target reward is large due to the exponential, which encourages the policy to move the foot closer to the center in the subsequent training iterations.


\paragraph{Progress Reward}
The target reward is a sparse reward, which is generally more difficult for DRL algorithms to optimize. 
We provide an additional dense progress reward to guide the character across the steps. 
More specifically, let $d_{t-1}$ and $d_t$ be the distance between the root of the character to the center of the desired step at the previous and the current time step, as projected onto the ground plane.
A progress reward 
$$r_{progress} = (d_{t-1} - d_t) \mathbin{/} dt$$
is added to encourage the characters to move closer to the stepping target.
$dt$ is the control period for each character in Table~\ref{tab:character-summary}.


\paragraph{Additional Reward For Humanoid}
It is common practice to incorporate task-agnostic rewards to encourage natural motion when working in the absence of any reference motion, e.g. \cite{duan2016rl_benchmark,Yu_2018_Symmetry}. 
We use similar reward terms to shape the motions for the Humanoid:
$$r_{addition} = r_\text{energy} + r_\text{limit} + r_\text{posture} + r_\text{speed} + r_\text{alive}.$$
The four terms penalize the character for using excess energy, reaching joint limits, failing to maintain an upright posture, and unnaturally speeding across the steps.
Most of the terms are adapted from the reward implementation in \cite{pybullet}.

For the energy penalty, we have
$$r_\text{energy} = -4.5\frac{1}{N_j}\sum_{j} \mid a_j \cdot v_j \mid - 0.225\frac{1}{N_j}\sum_{j} \mid a_j \mid^2,$$
where $N_j$ is the number of joints on the Humanoid, $a_j$ is the normalized torque for joint $j$, and $v_j$ is the joint velocity. 

The joint limit penalty is used to discourage the character from violating the joint limit, defined as  
$$r_\text{limit} =  -0.1\sum_j \mathds{1}_{j \notin [0.99l_j, 0.99u_j]}(j),$$
where $\mathds{1}_{j \notin [0.99l_j, 0.99u_j]}(j)$ is the indicator function for checking whether joint $j$ is beyond $99\%$ of its natural range of motion defined by limits $l_j$ and $u_j$.
In essence, this penalty is proportional to the number of joints near the lower or upper limit.

The posture penalty is 
$$r_\text{posture} = -\mid \alpha_x \mid \mathds{1}_{\alpha_x \notin [-0.4, 0.4]}(\alpha) - \mid \alpha_y \mid \mathds{1}_{\alpha_y \notin [-0.2, 0.4]}(\alpha),$$ where $\alpha_x$ and $\alpha_y$ are the roll and pitch of the body orientation in global frame.
The penalty applies only when the character is leaning sideways for more than 0.4 radians, or backwards beyond 0.2 radians or forwards by 0.4 radians.

We also observe that the Humanoid tends to move unnaturally fast to achieve a good progress reward. 
We add a velocity penalty 
$$r_\text{speed} = -\max(\norm{v_\text{root}}_2-1.6, 0),$$ 
to discourage the character from exceeding root speed of $1.6$ meters per second.
The issue does not effect Cassie since its speed is predetermined by the fixed gait period.

Finally, we add an alive bonus 
$$r_\text{alive} = 2$$ 
for every time step that the Humanoid is able to keep its root $0.7$ meters above the lower foot, otherwise the episode is terminated. 
This reward encourages the Humanoid to maintain balance and prevents it from being overly eager to maximize the progress reward.
\section{Learning Curricula}
\label{sec:curriculum}

The learning efficiency for the stepping stones task is strongly correlated to the distribution of the step parameters.
In this section, we describe five different sampling strategies, including uniform sampling for baseline comparison.
For clarity, we focus on the 2D parameter subspace of $(\psi, \theta)$, which are the relative yaw and pitch of consecutive steps shown in Figure~\ref{fig:step-generation}. However, we further extend this strategy to 3D and 5D
step parameter spaces.

\begin{figure*}%
  \centering
  \includegraphics[width=1.95\columnwidth]{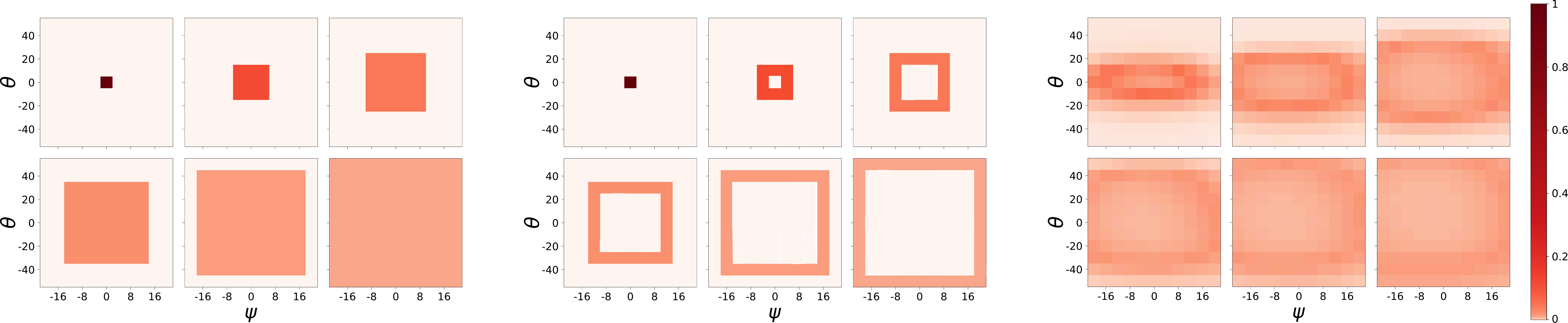}
  \caption{\textbf{Left}: The fixed-order curriculum advance evenly through the sampling space. \textbf{Middle}: The fixed-order boundary curriculum advance evenly, but only samples step on the boundary of the parameter space. \textbf{Right}: The adaptive curriculum is free to explore the parameter space at its own pace.}
  \label{fig:sampling_grid_all}
\end{figure*}

Except for uniform sampling, other strategies require dynamically adjusting the step parameter distributions.
As such, we first discretize the sampling space evenly into an $11 \times 11$ grid inside the region defined by $[\minus\psi_{\max}, \psi_{\max}] \times [\minus\theta_{\max}, \theta_{\max}]$. 
The midpoint of the grid is precisely $\psi = 0$ and $\theta=0$.
Also note that the granularity of the $\psi$-axis and $\theta$-axis is different since $\psi_{\max}$ is not equal to $\theta_{\max}$.
The discretization process is illustrated in Figure~\ref{fig:sampling_grid_all}.

\subsection{Uniform Sampling (Baseline)}
\label{sec:uniform-sampling}
The simplest strategy is to sample the parameters uniformly during training. 
This is effective if the sampling space only spans easy steps, e.g.~steps with small yaw and pitch variations.
As the step variations become larger, it becomes less likely for the policy to receive the step reward during random exploration, and so the gradient information is also reduced.
We also refer this strategy to as the no curriculum baseline, since it does not adjust step parameters distribution during training.

\subsection{Fixed-order Curriculum}
\label{sec:fixed-order-curriculum}
This curriculum is designed based on our intuition of tasks difficulty.
We first divide the $11 \times 11$ grid into six stages, from the easiest to the most challenging.
In stage~$k$, $\psi$ and $\theta$ are sampled uniformly from the $ (2k-1) \times (2k-1)$ grid centered at the middle point. 
E.g.~in the first stage, we only sample the center point of the grid, which means that every step is generated with $\psi = 0$ and $\theta = 0$. 
The curriculum advances when the average total reward during a training iteration exceeds a threshold (see Table~\ref{tab:curricula-summary}). 
The curriculum becomes equivalent to uniform sampling when the last stage is reached, i.e.~$k=6$, and is fixed until the end of the training. 
The process is illustrated in Figure~\ref{fig:sampling_grid_all}.
We call this the fixed-order curriculum because the stages proceed in a predefined order, although the progression from one stage to the next is still tied to the performance.
Similar approaches have been shown to be effective for learning locomotion tasks, e.g.~\cite{Yu_2018_Symmetry}.

\subsection{Fixed-order Boundary Curriculum}
This strategy is similar to the fix-order curriculum with one important modification: Instead of sampling uniformly in the 
rectangular domain, it only samples in the boundary regions.
Please refer to Fig~\ref{fig:sampling_grid_all} for visual illustration of the differences.
The premise is that the policy can remember solutions to previously encountered step parameters, or that the solution which solves the new parameters also solves the inner region, and so it is more efficient to sample only on the boundary.



\subsection{Difficult-tasks-favored Sampling}
\label{sec:difficult-curriculum}
This strategy is equivalent to the adaptive sampling introduced in \cite{Park_2019_siggraph_asia_learn_and_predict} and \cite{Won_2019_siggraph_asia_body_shape}. 
The idea is that during task sampling, more difficult tasks will cause more failure, leading to more frequent early termination. 
Because of this, even though the tasks are sampled uniformly, the data collected will be more biased towards easier tasks. 
To counter this, the sampling distribution is updated based on the current value function estimate of each task. 
This results in more difficult tasks being sampled more frequently, thus balancing the data distribution observed during training. 
In many ways, this strategy takes the opposite approach of the fixed-order curriculum, where the policy focuses on easy steps in early stages of training and moves progressively into more difficult settings.
We describe the implementation together in Section~\ref{sec:adaptive_curriculum}.



\subsection{Adaptive Curriculum}
\label{sec:adaptive_curriculum}
The motivating philosophy of our adaptive curriculum is that it is beneficial to avoid scenarios that are either too easy or too challenging during learning.
Most of the trajectory samples should be devoted to medium difficulty steps that the policy can improve on in the short term.

We define the capability of a policy $\pi$ for parameters $\psi$ and $\theta$ as 
$$C^\pi(\psi, \theta) =\int_s V^\pi(s, \delta_1(r, 0, 0), \delta_2(r, \psi, \theta))ds,$$ where $r$ is fixed to be $(r_{max}+r_{min})/2$ and $\delta_k(r, \psi, \theta)$ converts the step parameters to Cartesian vectors used by the policy and value function.
In simple terms, the capability metric is an answer to the question: Given two upcoming steps, what is the average performance of the current policy across all observed character states?

Evaluating $C^\pi$ is generally intractable, so we estimate it by executing the policy on an easy terrain, i.e.~the terrain generated by the first stage defined in Section~\ref{sec:curriculum}, once per episode. 
Each time the character makes contact with the target foothold, the curriculum evaluates the capability of the current policy for each $(\psi, \theta)$ pair in the grid by imagining their placements.
We refer to these step parameters as imagined since they are only used for computing the policy capability and do not reflect the actual placements of the physical footholds.
The process is repeated for five steps to accumulate different character states, and the mean result is used as a proxy for capability.
Also, note that only the parameters of the second step are used for evaluating the capability, i.e.~the first step is always fixed.
It is possible to use both steps for evaluation, but the second step will be replaced when the character makes contact with the first, since new steps are generated on every contact.
Lastly, we observe that the value function is less sensitive to the second step for Cassie, possibly due to the pre-trained imitation controller, and so we vary the first step instead. 

We then define the sampling probability of a set of parameters $(\psi_i, \theta_j)$ in the parameter grid to be proportional to 
$$f(\psi_i, \theta_j) = \exp(-k\mid \frac{C^\pi(\psi_i, \theta_j)}{C^\pi_{\max}} - \beta \mid),$$ 
where $C_{\max} = \max_{\psi, \theta}C^\pi$.
Finally, this proportionality is normalized into a probability distribution $p(\psi_i, \theta_j) = f(\psi_i, \theta_j) / \sum f(\psi, \theta)$. 
Here $k > 0$ controls the sensitivity to differences in capability values and $\beta \in [0, 1]$ decides the difficulty setting of the curriculum.
In our experiments, we use $k=10$ and $\beta = 0.9$ for the Humanoid and $\beta = 0.85$ for Cassie.

When $\beta = 1$, the curriculum prefers step parameters such that $C^\pi(\psi, \theta) = C^\pi_{\max}$, 
i.e.~steps where the policy has high confidence.
In practice, these usually correspond to the easiest steps, e.g.~ones without roll and pitch variations.
Conversely, if $\beta = 0$, the curriculum samples steps that are beyond the capability of the current policy. 
We use this as our implementation of difficult-tasks-favored sampling, as they are similar in spirit. 

\section{Results and Evaluations}
\label{sec:results}

\begin{table}
\centering
\caption{\label{tab:curricula-summary}Curriculum and learning parameters.}
\begin{tabular}{@{}lcrr@{}}
\toprule
Property & \phantom{abc} & Humanoid & Cassie\\
\midrule
\textit{Fixed-order reward threshold} && 2500 & 1000\\
\textit{Adaptive curriculum $\beta$} && 0.9 & 0.85\\
\textit{Exploration noise (logstd)} && $\minus1.5$  & $\minus2.5$\\
\textit{Samples per iteration ($\times10^4$)} && 5 & 4\\
\bottomrule
\end{tabular}
\end{table}

\begin{table*}

\begin{center}
\caption{\label{tab:policy_eval}Performance evaluation of policy under different settings, for Uniform (U), Fixed-Order (FO), Fixed-Order Boundary (FOB),
and Adaptive (A) curricula. 
The performance numbers represent the maximum radial distances $r$ achievable. 
Please see the text for a detailed explanation of the performance numbers.  Larger is better.
Bold indicates best compared to alternative. Entries marked with a dash indicate the policy fails for all $r \in [r_{\min}, r_{\max}]$.
}
\renewcommand{\arraystretch}{1.2}
\begin{tabular}{@{}lrrrrcrrrr@{}} 
\toprule
& \multicolumn{4}{c}{Humanoid} & \phantom{abc}& \multicolumn{4}{c}{Cassie}\\
\cmidrule{2-5}
\cmidrule{7-10}
\textit{Task Parameter} & \textit{U} & \textit{FO} & \textit{FOB} & \textit{A} && \textit{U} & \textit{FO} & \textit{FOB} & \textit{A} \\
\midrule
\textit{Flat} $(\theta = 0)$\\
\qquad $\psi = 0$ & 1.20, 1.20 &1.20, 1.25 &1.35, 1.35 & \textbf{1.45}, \textbf{1.50} && 0.85 & 0.90 & \textbf{0.95} &\textbf{0.95}\\
\qquad $\psi = 20$ &1.15, 1.20 &1.15, 1.20 &1.25, 1.35 &\textbf{1.35}, \textbf{1.40} && 0.75 & 0.80 & 0.85 & \textbf{0.90}\\
\textit{Single-step $(\psi = 0)$}\\
\qquad $\theta = 50$ & --- &0.75, \textbf{0.80} & --- &\textbf{0.80}, \textbf{0.80} && \textbf{0.80} & 0.80 & \textbf{0.85} &0.60\\
\qquad $\theta = \minus 50$ & 1.30, \textbf{1.50} & \textbf{1.50}, \textbf{1.50} &0.75, 1.00 & 0.90, 0.95 && 0.80 & \textbf{0.85} & 0.80  & 0.75\\
\textit{Continuous-step $(\psi = 0)$}\\
\qquad $\theta = 50$ & --- & --- & --- & ---, \textbf{0.65} && --- & 0.40& \textbf{0.45} & 0.40\\
\qquad $\theta = \minus 50$ & --- &\textbf{0.75}, \textbf{0.80} &---, 0.65 &0.65, 0.70 && --- & --- & --- & \textbf{0.35}\\
\textit{Spiral $(\psi = 20)$}\\
\qquad $\theta = 30$ & --- &0.75, 0.80 & --- &\textbf{0.80}, \textbf{0.85} && --- &0.50 & \textbf{0.65} & 0.60\\
\qquad $\theta = \minus 30$ & 0.65, 0.70 &\textbf{1.40}, \textbf{1.50} & 0.65, 0.75 & 1.00, 1.10 && --- &0.55 & --- & \textbf{0.60}\\
\bottomrule
\end{tabular}
\end{center}
\end{table*}

We train stepping stone policies for the Humanoid, Cassie and Monster. 
We then quantitatively evaluate and compare the differences between sampling strategies. 
Since the Humanoid and the Monster are similar in terms of control and reward specifications, 
we focus our evaluation on the Humanoid and Cassie.

\subsection{Summary}
We first summarize the high-level findings.
All three curricula that gradually increase the task difficulty are able to do well at solving the stepping stone tasks.
This include the fixed-order, fixed-order boundary, and adaptive curricula.
The remaining approaches, uniform sampling and difficult-tasks-favored sampling, each produce conservative policies that simply learn to stand on the first step when the alive bonus is present, and otherwise yield much less robust and less capable policies.
The performance of the policies is best demonstrated in the supplementary video.

The differences between fixed-order, fixed-order boundary, and adaptive curricula is small in terms of learning speed and control performance.
One practical advantage of fixed-order curricula is that they are intuitive and simple to implement.
The computational overhead for implementing these curricula is negligible compared to simulation and controller inference.
In comparison, the adaptive curriculum is more flexible in two aspects. 
First, it uses implicit parameter $\beta$ to adjust the difficulty of the task, rather than task-dependent reward thresholds.
Additionally, it allows for more fine-grained curriculum control for independent task parameters, such as yaw and pitch in our experiments.
In summary, we suggest adaptive curriculum be used for tasks for which the difficulty cannot be intuitive assessed, especially when multiple tasks parameters are involved.
For other tasks, fixed-order curricula may be more suitable due to their simplicity.

\subsection{Policy Structures}
All policies in our experiments are represented by two five-layer neural networks, 
each hidden layer has $256$ neurons, and trained with PPO. 
One network is the actor that outputs the mean of a Gaussian policy and the other is the critic that outputs a single value which indicates the value function estimate of the current policy. 
The first three hidden layers of the actor use the softsign \cite{turian2009softsign} activation while the final two layers use ReLU activation. 
We apply the hyperbolic tangent function (Tanh) to the final output to normalize the action to have a maximum value of one.
For the critic, we use ReLU for all the hidden layers.
The policy parameters are updated using Adam optimizer \cite{kingma2014adam} with a mini-batch size of $1024$ and a learning rate of $3\times 10^{\minus5}$ for $10$ epochs in each roll-out.
The learning pipeline is implemented in PyTorch \cite{paszke2019pytorch}. 
For fixed-order curricula, training a single policy takes about $12$ to $24$ hours on a GPU, with simulation running in parallel on a 16-core CPU.
For the adaptive curriculum, the training time depends heavily on the dimensionality of the step parameter space, since the capability is evaluated prior to each roll-out.
In the 5D parameter case, the value function is evaluated $5 \times 11^5 \approx 8 \times 10^5$ times for each policy update.
However, this computation can be done in parallel using batched evaluations on a GPU.

To reduce the amount of computation, we pre-train an initial, straight line and flat terrain, locomotion controller for both the humanoid and Cassie.
The step length $r$ is sampled from $[0.65, 0.8]$ for the humanoid and $[0.35, 0.45]$ for Cassie.
These controllers are used as the starting point for all subsequent experiments.
This also means that we are directly comparing different sampling strategies on their performance for the stepping stones task.
For the experiments described in this section, we use $\psi_{\max} = 20^\circ$ and $\theta_{\max} = 50^\circ$ unless otherwise specified. 
Other character-specific curriculum and learning parameters used for training are summarized in Table~\ref{tab:curricula-summary}.

\subsection{Learning Curves for 2D Parameter Space}
\begin{figure}
  \centering
  \includegraphics[width=0.98\columnwidth]{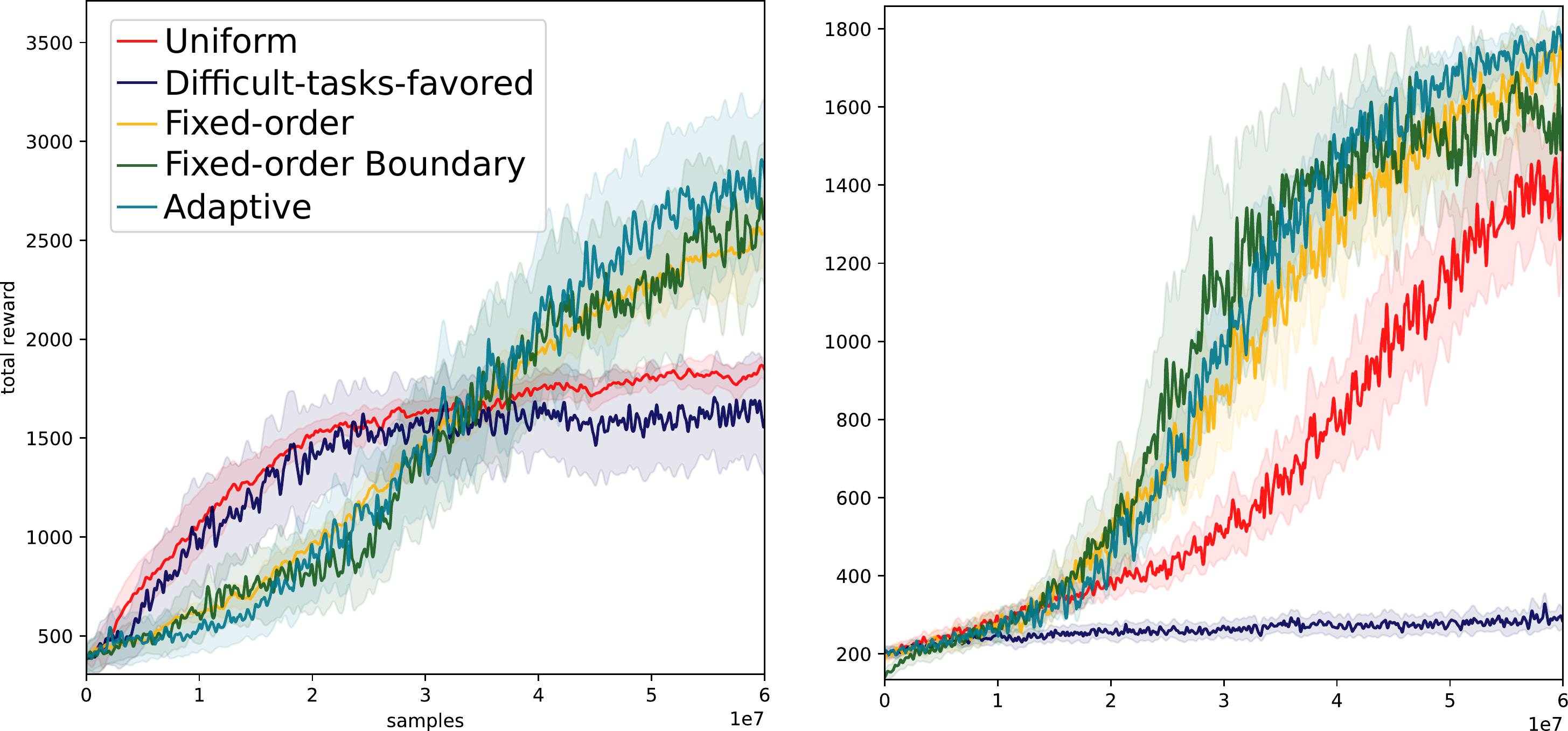}
  \caption{Learning curves for different sampling strategies, averaged over five runs. \textbf{Left}: Humanoid. \textbf{Right}: Cassie.}
  \label{fig:learning_curve}
\end{figure}

The performance of different sampling strategies is shown in Figure~\ref{fig:learning_curve}.
To ensure fairness in the learning curves comparison, we use uniform sampling to evaluate all policies.
It is important to note that the learning curves may not reflect the performance of the policies as precisely as visual demonstrations.
In particular, due to the presence of the alive bonus for the Humanoid, a simple policy can receive a maximum reward of 2000 by standing still on the first step.
Please refer to the supplementary video for further details.

For the Humanoid, the learning curves capture the phenomenon of local and global optima, where the sampling strategies fall into two categories.
In the first category, both the uniform and difficult-tasks-favored sampling strategies quickly achieve decent performances, but eventually converge to lower final rewards.
The combination of difficult steps and sparse target reward discourages the policies trained with these two methods to make further progress after learning to balance on the first step.
In contrast, the policies steadily improve under the fixed-order, fixed-order boundary, and adaptive curricula, due to the gradual build-up of steps difficulty. 
These three curricula were able to guide the policies to solve the stepping stones task, and the difference in learning speed is insignificant.
The distinction between these three curricula is more clear in their use cases, which we discuss in the next section.

\subsection{Curriculum Progress for 2D Parameter Space}
The fixed-order curriculum is developed based on our intuition of task difficulty.
However, the relationship between a task parameter and difficulty is not always obvious.
The benefits of the adaptive curriculum are that it yields a smoothly-advancing curriculum with 
fine-grained step distribution control based on the policy's local capability. 

Figure~\ref{fig:sampling_grid_all} shows the relative progress of the fixed-order and adaptive curriculum, where the heatmaps of the latter were captured at the end of each of the six stages.
From the adaptive curriculum heatmaps, it is clear that the competency in the yaw dimension expands much faster than in the pitch dimension. 
This observation is consistent with our intuition that variations in the yaw dimension should be easier to learn.
Furthermore, the high-probability, ring-structured region of each heatmap resembles that in the fixed-order boundary curriculum.
Overall, the adaptive curriculum is flexible and has similar features to the fixed-order and fixed-order boundary curricula.
One disadvantage is that it requires more computation to evaluate the capability of the policy.

\subsection{3D Parameter Space}
We extend the evaluation of fixed-order, fixed-order boundary, and adaptive curriculum to the 3D parameter space, now including step distance $r$. 
The step distance is sampled from 11 uniformly discretized values between $[0.65, 1.5]$ meters for the Humanoid and $[0.35, 1.0]$ for Cassie. 
For the fixed-order curriculum, in addition to the parameters defined Section~\ref{sec:fixed-order-curriculum}, it starts at $r = r_{\min}$ in the first stage and expands the sampling space by two grid points every time the reward threshold is met. 
The fixed-order boundary curriculum is similarly extended.
For the adaptive curriculum, the capability of the policy defined in Section~\ref{sec:adaptive_curriculum} is modified to take an additional parameter $r$.

For the fixed-order curriculum, it may be impossible to progress to the final stage due to the physical capability of the characters. 
However, it is entirely possible that a parameter choice, e..g,  $(r = r_{\max}, \psi = 0, \theta = 0)$, 
is within capability limit, and that the fixed-order curriculum will never have the chance to attempt it, 
while the adaptive curriculum is free to advance unevenly in the parameter space. 
We observe this phenomenon in our experiments.

\begin{figure*}
  \centering
  \includegraphics[width=0.9\linewidth]{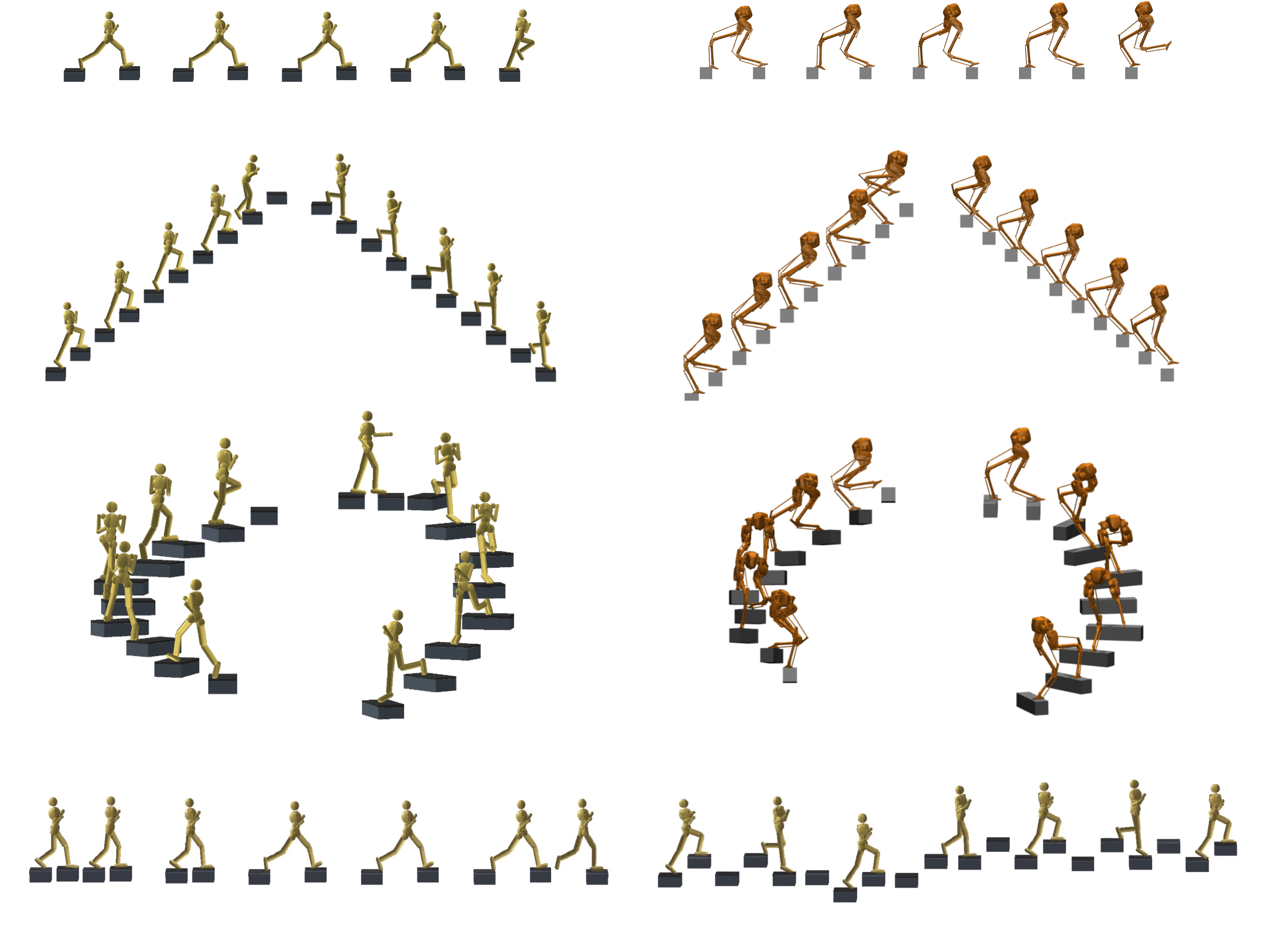}
  \caption{Snapshot of the motions on different test scenarios.}
  \label{fig:3d-scenarios}
\end{figure*}

\paragraph{Policy Capability Limits}
We also examine the performance of the policies by fixing $\psi$ and  $\theta$ while pushing $r$ to the limit.
The test scenarios are summarized in Table~\ref{tab:policy_eval}.
The single-step scenario means one inclined or declined step at the start, followed by horizontal straight-line steps until the end.
The continuous-step variation is where all steps are on a constant incline or decline.
The motions for some of the scenarios can be visualized in Figure~\ref{fig:3d-scenarios}. 
\begin{figure}
  \centering
  \includegraphics[width=0.9\columnwidth]{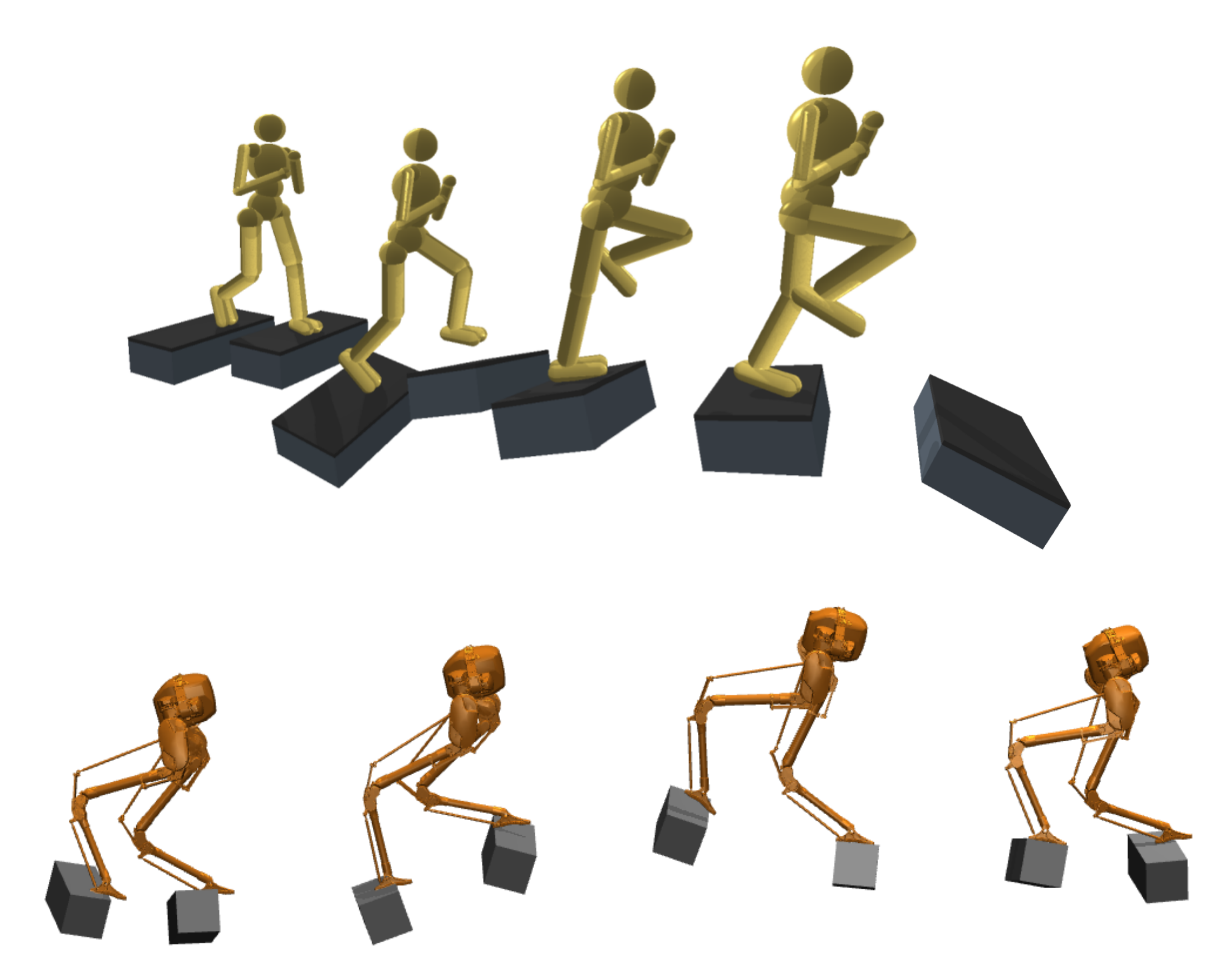}
  \caption{Steps with roll and pitch variations.}
  \label{fig:tilted-5d-steps}
\end{figure}

We test whether the policy can sustain the performance level for ten consecutive steps. 
For the Humanoid, the simulation is not fully deterministic due to an observed underlying stochasticity in PyBullet's contact-handling, 
and so we repeat each scenario five times and record two numbers. 
The first represents the maximum value of $r$ for which the policy succeeds for all five runs, and it thus provides a conservative estimate.
The second number represents the maximum value of $r$ for which the policy succeeds in at least one of the runs.
We observe empirically that the policies work consistently when $r$ is less than the maximum value recorded, and thus the learned policies are generally quite robust.

When we change $\theta$ to $\minus 40$ degrees in the single-step and continuous-step decline scenarios for the Humanoid, the adaptive curriculum is able to perform consistently for all five runs at 1.5 meters and 0.8 meters respectively.
This suggests that $\theta = \minus 50$ may be near the physical limit of the Humanoid.
Since the adaptive curriculum prioritizes medium difficulty settings, e.g.~$\beta=0.9$, the most extreme scenarios are likely to be sampled very rarely.
The fixed-order curriculum does not suffer from this issue since it is forced to sample the extreme scenarios as long as the final stage is reached.

\subsection{5D Parameter Space}
For the 5D parameter space, we also include the pitch and roll of each step, as measured in their respective local frames, so that the generated steps are tilted. 
We sample $\phi_{x}, \phi_{y} \in [\minus20, 20]$ degrees, where $\phi_x$ and $\phi_y$ are the roll and pitch of the steps. 
Each new dimension is discretized into $11$ intervals as before and the adaptive curriculum is applied to train a new policy for each character.
For comparison with their respective 3D policy, we evaluate the number of steps each policy can handle on ten randomly sampled 5D stepping stone sequences, each with 50 steps.
The mean and standard deviation of successful steps is reported in Table~\ref{tab:5d-robustness}.
In general, the 5D policies are significantly more robust than their 3D counterparts, since they are aware of the roll and pitch variations of the individual steps.
A snapshot of the motion on tilted steps can be seen in Figure~\ref{fig:tilted-5d-steps}.

\begin{table}[h]
  \centering
  \caption{\label{tab:5d-robustness}Robustness of 3D and 5D policies on 5D stepping stone   sequences.  The numbers represent the number of steps before falling.}
  \begin{tabular}{@{}lcrcr@{}}
  \toprule
  Parameters && Humanoid && Cassie\\
  \midrule
  \textit{3D Policy} && $25.9 \pm 14.6$ && $22.0 \pm 15.8$\\
  \textit{5D Policy} && $50 \pm 0$ && $45.7 \pm 8.6$\\
  \bottomrule
  \end{tabular}
\end{table}

\begin{figure}
  \centering
  \includegraphics[width=0.85\columnwidth]{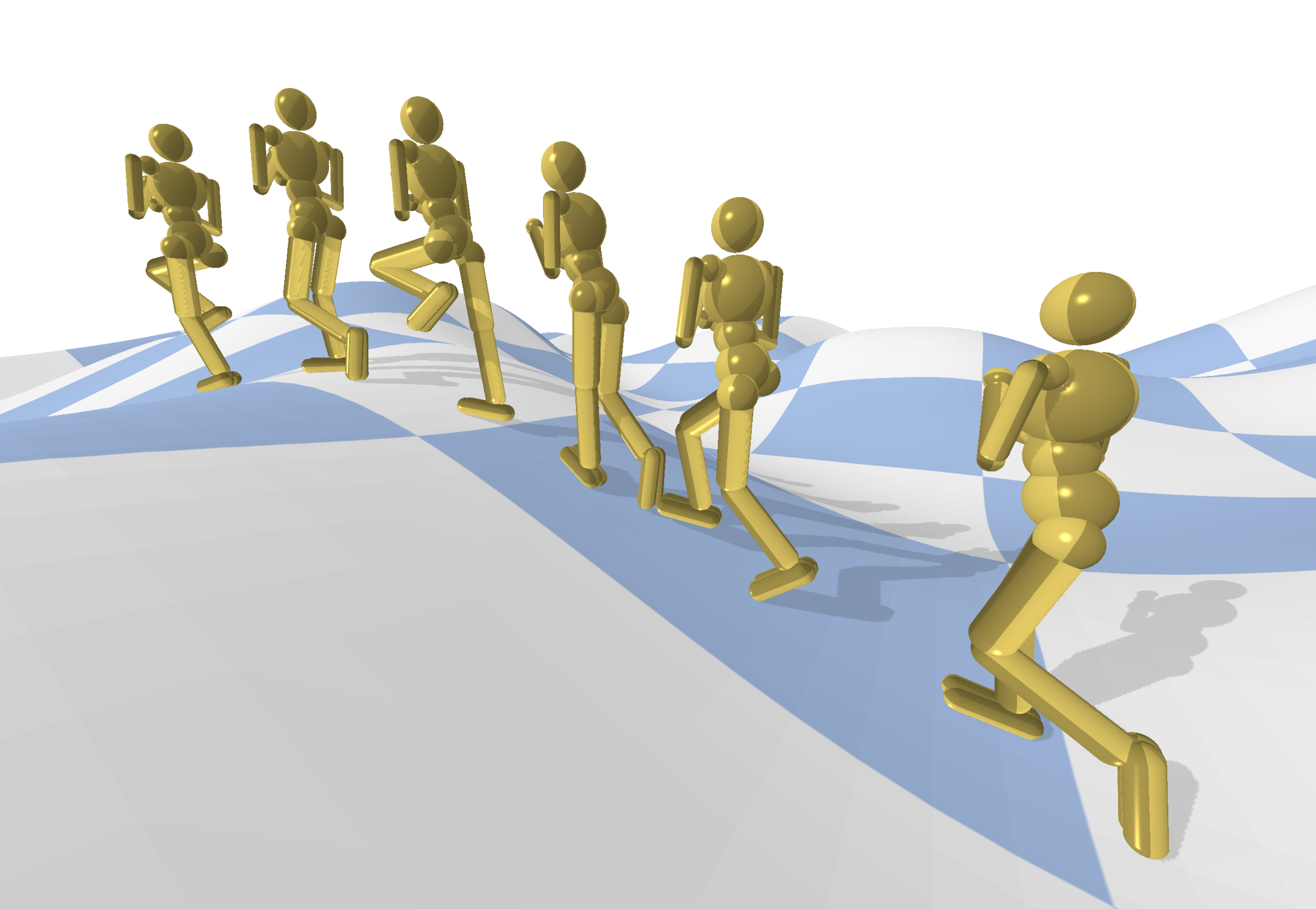}
  \caption{Stepping-stone policy applied to continuous terrain.}
  \label{fig:humanoid-terrain}
\end{figure}

\subsection{Walking on Variable Terrain}
Given the considerable abilities of the characters to realize challenging stepping stone scenarios, we expect that the same control policies can execute similar steps on continuous terrain as it does on isolated footholds. 
The primary difference between the two scenarios is that the continuous terrain
might present tripping hazards for the swing foot that are not present in the case of isolated stepping stones.
Also, the continuous terrain may demand more precise foot placements since the surfaces near target locations have non-uniform slopes.
We use the height field primitive in PyBullet to model continuous terrains generated using Perlin noise.
We first synthesize a footstep trajectory on the xy-plane starting from the character's initial position, and turning five degrees every step to ensure that the entire trajectory remains on the height field.
Then we project the footstep trajectory onto the height field to create 5D stepping stone sequences for the control policies.
Note that the policy perceives discrete steps, as before, while the simulator sees only the height field.
While we find height fields in PyBullet to have slightly different contact dynamics than the discrete footholds we used for stepping stones, our policies are robust enough to handle the differences without further training.
Figure~\ref{fig:humanoid-terrain} shows the Humanoid walking on continuous terrain.

To demonstrate the generality of our approach, we apply the same learning pipeline to train a policy for the Monster with the same 5D parameter space. 
This policy achieves the same robustness and capabilities on the continuous terrain.
Please refer to the supplementary video for visual results.
For the Monster character, the control policy finds gaits that perform tiptoe walking.
We attribute this to several possible reasons.
First, the character weighs comparatively less than the Humanoid, thereby making this strategy a feasible option.
Additionally, the alive bonus (\S\,\ref{sec:rewards}) encourages the policy to maintain a minimum torso height of 0.7 meters 
relative to the lower foot.
A toe walking gait emerges as the benefit of the increased height margin outweighs the cost of the increased energy expenditure.
Lastly, the final behavior of the policy may show some variations across different runs, due to the stochastic nature of the policy initialization and the optimization.

\section{Discussion and Limitations}
\label{sec:discussion}

During training, we use stepping stone blocks which are five times wider than the ones used for rendering.
We find this to improve the training consistency, as it makes the sparse target reward more discoverable during random exploration.
However, it also causes the characters to occasionally miss the step for some extreme sampling parameters when testing on 
narrower steps. 
This issue could be addressed by adding step width as a curriculum parameter and decrease it over time during training.

The Humanoid and Cassie appear to use different anticipation horizons.
Although we provide a two-step look-ahead for both the Humanoid and Cassie, the value function estimates indicate that Cassie's policy considers only the first step while the Humanoid uses both. 
This may be because Cassie has a fixed step-timing, enforced by the phase variable, which limits the policy to take more cautious step. 
For the Humanoid, we observe that its step-timing depends on the combinations of the two upcoming steps. 
For example, the character prefers to quickly walk down consecutive descending steps, while taking other combinations more slowly. 
This gives the policy more flexibility and makes the second step information more meaningful.

For the adaptive curriculum, we estimate the difficulty of a step by imagining it while traversing horizontal and straight steps. 
One limitation of this method is that it ignores the influence of step transitions. 
For example, it is generally easier to make a right-turning step if the swing foot is the right foot, and vice versa.
A natural way to take the transition into account is to estimate the difficulty of the step before the step generation within the training episode. 
However, this requires additional computation.

The purpose of look-ahead delay was to emulate human reaction time to produce more conservative motions.
With the default delay of 30 frames, the Humanoid walks across the stepping stones at an average speed of 1.35~m/s, similar to typical human walking pace. 
We can control the walking speed by adjusting the look-ahead delay and disabling the speed penalty. 
When the look-ahead delay is set to 2 frames, the Humanoid traverses the terrain at an average speed of 2.10~m/s, which is closer to jogging.

Despite being able to control arm movements, the Humanoid prefers to maintain a tucked position 
for its arms, as a result of the energy penalty. 
The Humanoid also may briefly stub its toes on the upcoming step when climbing up steep step sequences. 
To further improve the motion quality of the Humanoid, it should be possible to 
pre-train a locomotion controller using motion capture data. 
This pre-trained controller should then retain the basic motion style 
when it is subsequently optimized for the stepping stones task, as was the case for the Cassie results.
Further stylistic rewards may be helpful for fine-tuning the motion quality.

Lastly, our policies come close to reaching the physical limits achievable with a normal stepping gait.
Alternate locomotion modes are required to solve even more drastic terrain variations, 
e.g., the Humanoid can use hands to clamber up steeper inclines.
Addressing the clambering problem may also pave way to extending our approach to learning stepping stone skills for quadruped characters.
\section{Conclusions}

We have presented a general learned solution capable of solving challenging stepping stone sequences, as applicable to physics-based legged locomotion. 
To this end, we evaluated four different curricula and demonstrated that the key to solving this problem is using suitable learning curricula that gradually increase the task difficulty according to the capability of the policy.
In the future we wish to integrate these stepping capabilities with a step planner, to rapidly generalize the capabilities to new characters, to support true omni-directional stepping, to integrate hands-assisted locomotion modes such as clambering, and to test the capabilities on physical robots.
We believe that the simplicity of our key findings, in retrospect, makes them the perfect \textit{stepping-stone} to future research on generalized locomotion.

\bibliographystyle{eg-alpha-doi}  
\bibliography{main}
\end{document}